# Energy Efficient Adaptive Network Coding Schemes for Satellite Communications

Ala Eddine Gharsellaoui[1], Samah A. M. Ghanem[2], Daniele Tarchi[1], and Alessandro Vanelli Coralli[1]

[1] Department of Electrical, Electronic and Information Engineering,
University of Bologna, Bologna, Italy
[2] Huawei R&D Labs, Stockholm, Sweden

**Abstract.** In this paper, we propose novel energy efficient adaptive network coding and modulation schemes for time variant channels. We evaluate such schemes under a realistic channel model for open area environments and Geostationary Earth Orbit (GEO) satellites. Compared to non-adaptive network coding and adaptive rate efficient network-coded schemes for time variant channels, we show that our proposed schemes, through physical layer awareness can be designed to transmit only if a target quality of service (QoS) is achieved. As a result, such schemes can provide remarkable energy savings.

**Key words:** Energy Efficiency, Network Coding, Satellite Communications

## 1 Introduction

Network coding is a transmission technique that, by performing algebraic operations across transmitted packets rather than relying on packet repetition or replication, allows to reliably transmit with lower end to end delays in a communication system. Additionally, network coding mechanisms are key enablers to energy efficient communications. Due to the steady increase in energy consumption and energy costs in mobile communication systems, more efficient schemes are required. In particular, with higher reliability obtained via network coding, less re-transmissions are required. Consequently, more energy savings can be achieved [1]. Moreover, when the network coded schemes are specifically designed for enhancing their awareness with respect to the system characteristics, higher performance gains can be achieved in terms of delay, throughput or energy efficiency [2, 3]. One of the most important issue to be considered in satellite communications is energy efficiency. In [4], several factors that have a direct impact on energy efficiency of satellite and mobile terminals have been discussed, including dynamic spectrum access and cross layer design. In [2] and [5], the authors show that channel-aware transmission schemes jointly with network coding, can serve to reduce the delay and allow for energy performance gains. In [3], the authors propose novel adaptive network coding schemes and show a clear trade-off between energy-driven channel-aware schemes, that remain silent when channel encounter high erasures, and rate-driven channel-aware schemes that chose to transmit more to account for erasures.

In this work, various aspects of energy efficiency using network coding and modulation schemes are proposed. The schemes are evaluated in a realistic satellite channel model for open area environments and Geostationary Earth Orbit (GEO) satellites. Simulation results demonstrate clear trade-offs among average number of transmissions, delay, throughput and energy efficiency. We highlight that adaptation through channel-aware policies allows for silence periods or less transmissions which leads to significant energy savings compared to non-adaptive network coding or adaptive network coded schemes that are rate efficient.

## 2 System Model

Our focus is on a GEO satellite communications system forward link transmission, by considering a mobile terminal with constant speed in a open area environment. In such system, the transmitter performs a Random Linear Network Coding (RLNC), which is a Network Coding (NC) scheme that relies on coding across the packets using random linear coefficients in order to increase the transmission reliability mimicking the wireless diversity concept. The open area environment is modeled by resorting to the Land Mobile Satellite (LMS) model in [6], that is one of the most known in the literature. This model is based on a joint exploitation of a state based and a Loo based distribution [7] that allow to efficiently reproduce the shadowing and fading effects of a forward link satellite channel under mobile terminal assumptions. In this paper, we capitalize on the coded/uncoded packet transmission over the Markov model proposed in [2] to analyze our proposed schemes that rely mainly on channel variation awareness. Each state in the Markov model is represented by the couple $(i, h_j)$ that stands for the number $i$ of coded/uncoded packets to be sent and the channel state $h_j$, whose value varies over time. Therefore, such Markov model can be expressed by a one-step transition probability matrix $P$, whose size is defined by a finite number of states, and its components are defined by using two transition probability components: $p_{(i,h_j)\to(i-1,h_{j+1})} = 1 - P_e(h_j)$, and $p_{(i,h_j)\to(i,h_{j+1})} = P_e(h_j)$, where $P_e(h_j)$ is the packet erasure probability at the channel state $h_j$ for the duration of the packet transmission, and the probability of transitioning from the channel state and back to itself equals zero due to channel variation over time. This means that for a packet of size B bits, and with bit error probability $P_b(h_j)$ at a given channel state $h_j$, the erasure probability is given as, $P_e(h_j) = 1 - (1 - P_b(h_j))^B$. We resort to the approximation of the delay under network-coded transmission provided in [2], where the expected time to deliver $N_i$ coded packets is given as:

$$T(i, h_j) = T_d(N_i, h_j) + \sum_{l=1}^{i} P^{N_i}_{(i,h_j)\to(l,h_{j+N_i})} \; T(l, h_{j+N_i+1}), \qquad (1)$$

with $T_d(N_i, h_j) = N_i T_p + T_w$, where $T_p$ is the duration of one packet, and $T_w$ is the waiting time for acknowledgment. $P^{N_i}_{(i,h_j)\to(l,h_{j+N_i})}$, corresponds to the transition probability between states at the $N_i^{th}$ transition of matrix $P$. Finally, in $j + N_i + 1$ the term $+1$ appears due to acknowledgment. Fig. 1 illustrates the Markov chain as proposed in [2]. This model of coded and uncoded packet transmission over time varying channels assumes a finite number of time slots for sending a given number of packets. Thus, the model and the delay approximation inherently constraints the number of re-transmissions of packets, but has sufficiently large number of slots for a reliable approximation.

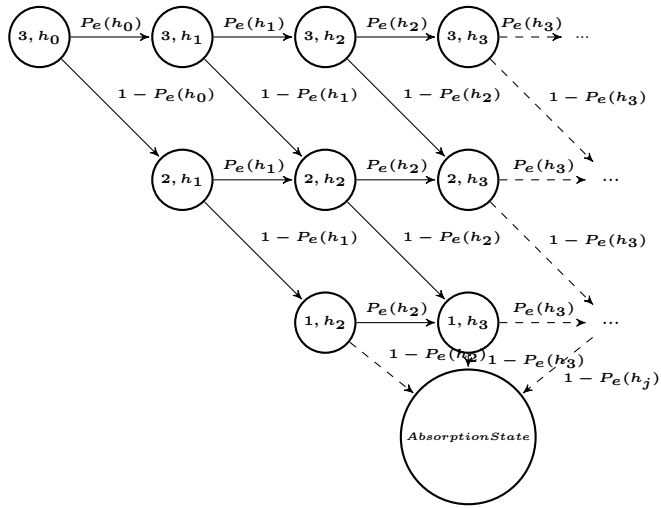

**Fig. 1.** Time Varying Channel Model of 3 Packets Transmission in [2]

## 3 Energy Efficient Adaptive Network Coding Schemes

The main objective is to propose energy efficient schemes by the exploitation of the adaptation of the coded packets transmission to the channel awareness under different levels of algorithm complexities. We discuss three proposed schemes and compare them to non-adaptive network coding scheme for time varying channels and to the rate driven adaptive network coded scheme in [2].

### 3.1 Adaptive Network Coding with Energy Efficiency (ANCEF)

This scheme adapts the transmission for achieving the energy efficient, by following the observation of the channel erasure; the strategy is to transmit small batches of coded packets if the observation of channel erasure is high (applies to low SNR), and to transmit larger batches of coded packets if erasure is less (applies to high SNR). Through this, the system can reduce transmissions and re-transmissions and save energy. Furthermore, QoS measure has been introduced to design such algorithm. In particular, if a bit error probability $P_b$ less than $10^{-5}$ is not met[1], the transmitter will choose to be silent with no transmissions. Therefore, much energy savings can be obtained. The following equation illustrates the number of coded packets $N_i$ required to be sent at the channel state $h_j$ when $i$ degrees of freedom (dof)[2] are required at the receiver,

$$N_i = \sum_{s=j}^{j+i-1} (1 - P_e(h_s)). \qquad (2)$$

It's worth mentioning that $N_i$ required to be rounded to nearest decimal or integer number, because it represents the number of coded packets. Moreover, it is worth to note that the sum is expressed with a shifted start of the state of measurements. This is due to the channel evolution over time, thus, a new round of transmission/re-transmission is associated to shift in the channel window. When erasures are high (at low SNR) such sum vanishes to zero corresponding to no transmission. However, when erasures are very low (at high SNR) such sum converges to the transmission of $i$ degrees of freedom almost surely.

### 3.2 Self-Tracing Adaptive Network Coding with Energy Efficiency (STANCEF)

In this scheme, we propose a self-tracing ANCEF scheme, which improves ANCEF by adding to the observation of the channel erasure the capability of looking-forward into the channel erasures if, looking- backwards, the packets at earlier transmissions are lower than the dof. Thus, the transmission strategy of such algorithm is similar to the ANCEF discussed in previous section, where less coded packets will be transmitted adaptively at high erasures, and more packets will be transmitted adaptively at low erasures. However, there is an amount of coded packets $\Delta_i$ needed to be as additional amount of future re-transmissions to establish all lost dof. Therefore, such $\Delta_i$ decreases as we move towards higher SNR, until it vanishes to zero when the transmission strategy of $N_i = i$. Once again, a certain QoS measure needs to be met, otherwise, the transmitter remains silent. The following equation represents the STANCEF transmission strategy of the number of coded packets $N_i$ required to be sent at a certain channel state $h_j$:

$$N_i = \sum_{s=j}^{j+i+\Delta_i-1} (1 - P_e(h_s)), \qquad (3)$$

---

[1] The acceptable bit error rate acceptable by the ITU ranges between $10^{-3}$ to $10^{-6}$ based on the rate and service expected at the mobile terminal

[2] A degree of freedom corresponds to the number of linear combinations that are required at the receiver to allow decoding the RLNC combined packets

where $\Delta_i$ is the foreseen losses due to self-tracing which is the difference between $i$ dof required at preliminary transmission at state $h_{j*}$ and the number of coded packets $N_i$ adapted to the channel at the same state. It is given by, $\Delta_i = i - N_i$, where $\Delta_i$ equals 0 at the initial state of first transmission. However, $\Delta_i$ is higher or equal to 0, at $h_j^*$ corresponds to zero or more coded packets contributed at a re-transmission stage one step ahead of its previous transmission. Thus, $\Delta_i$ at $h_j^*$ will contribute to $N_i$ at a forward channel state $h_{j*+N_{i+1}+1}$, where the addition of one represents the one step ahead due to ACK after preliminary transmission and before re-transmission.

### 3.3 Adaptive Network Coding and Modulation with Energy Efficiency (ANCMEF)

In this scheme, we integrated adaptive modulation to the ANCEF scheme. The rationale behind this, is that, on the one hand, a higher modulation order $m$ allows for transmitting the same amount of information in shorter packets due to the concatenation of more bits in the real and imaginary spaces. On the other hand, a higher modulation order is associated with higher energy per symbol, and less energy per bit, i.e. $E_b/N_0 = E_s/(N_0 * \log m)$. Thus, a higher bit error probability suggests that higher number of packets need to be sent due to adaptation. Such trade-off between the packet length and the number of coded packets for a given modulation scheme is of particular interest to address when taking into account energy efficiency. ANCMEF transmission strategy of coded packets $N_{i_m}$ is given by,

$$N_{i_m} = \sum_{s=j}^{j+i*\log m - 1} (1 - P_{e_m}(h_s)), \tag{4}$$

Resorting to the energy efficiency of the proposed scheme, the lower bound on random linear network coding, with $N_{i_m} \geq i$, i.e. with equality, is used, this was reflected in the sum range, by scaling the degrees of freedom $i$ by a factor $\log m$ that unifies the energy per symbol for each modulation scheme. $P_{e_m}(h_s)$ is the erasure probability of that modulation which can be derived as $P_{e_m} = 1 - (1 - P_{b_m})^B$, where $P_{b_m}$ is the bit error probability for the given modulation order $m$, and $B$ is the number of bits per packet.

Indeed the aim of the scheme is to find the optimal number of coded packets $N_i^m$ for a given modulation order $m$ to be transmitted/re-transmitted for assuring successful reception of a given number of $i$ dof along the way with energy efficiency; hence when $P_{b_m}$ of $m$-th modulation order is derived, for fair energy comparison among the different modulation schemes, $E_s$ is supposed to be constant for each modulation scheme.

## 4 Numerical Results

We shall now provide a set of illustrative results that cast further insights to the proposed schemes. Particularly, we focus on a satellite scenario and its related

LMS channel considering GEO satellite with delay $T_w$ equals 0.2388 $sec$, open area, and a mobile terminal with speed of 10 $m/s$. To construct the simulation environment, we consider to transmit 4 coded packets/dof, the maximum batch length due to channel adaptation of the schemes is constrained to 16 coded packets/dof, and the number of transmission/re-transmission trials is constrained to 10. This corresponds to a transition matrix of maximum size equals 401 × 401 including one additional slot for absorption state, which changes entries and size according to the transmitted packets.

The performance of the proposed schemes is compared in terms of average number of sent packets, delay, throughput, and energy efficiency.

The two benchmark schemes are the non-adaptive network coding scheme for time variant channels and the adaptive rate-efficient network coding scheme, both schemes are in [2]. In the non-adaptive network coding scheme, it is clear that the number of coded packets are fixed along the transmission/retransmission with no adaptation considered. The rate efficient adaptive network coding scheme is self explanatory, as it favors reliable transmission/reception and higher rates over energy efficiency. Contrary to the non-adaptive network coding scheme, the three proposed schemes adapt the number of coded packets for each batch of transmission based on the missing dof and on the channel erasures at a given window of estimated channel.

The modulation scheme considered is the BPSK in case of NC, ANC, ANCEF and STANCEF, while the ANCMEF exploits four possible modulation schemes, i.e., BPSK, QPSK, 8PSK and 16QAM, in order to efficiently exploit the channel behavior. The selection of modulation order is driven by the $E_s/N_0$ level.

The number of bits per packet is considered to be 1000, and since the maximum number of packets per batch equals 16, the maximum possible batch of packet size equals 16000 bits. This number of bits corresponds to the same, the half, the one-third, and the one-forth number of samples in BPSK, QPSK, 8PSK, and 16QAM, respectively.

First, the average number of coded packets sent back to back for the different schemes is compared; under different $E_s/N_0$ values and with mobility speed equals 10 $m/s$, this can be seen in Fig. 2. In general, we can see that the average number of packets for all the schemes at low SNR is greater than those at high SNR due to the higher probability of re-transmission at low SNR. Due to energy efficiency rule, the average number of packets will be as much small as to commensurate with the higher erasure probability at low $E_s/N_0$ values, since they are designed to limit the number of sent packets in case of bad channel conditions i.e., there is no need to spend more energy in such low chance of delivery. This can be emphasized looking into the low SNR, where the erasure probability is high such that energy efficient schemes favor not to send anything in order to avoid energy wasting; this is the contrary for the ANC benchmark that has been designed to achieve the highest possible rates. It is worth to note that the average number of coded packets in ANCMEF gets larger for higher $E_s/N_0$ values; this is expected since we aim to adapt the number of sent packets for achieving target reliability, keeping into account the energy efficiency.

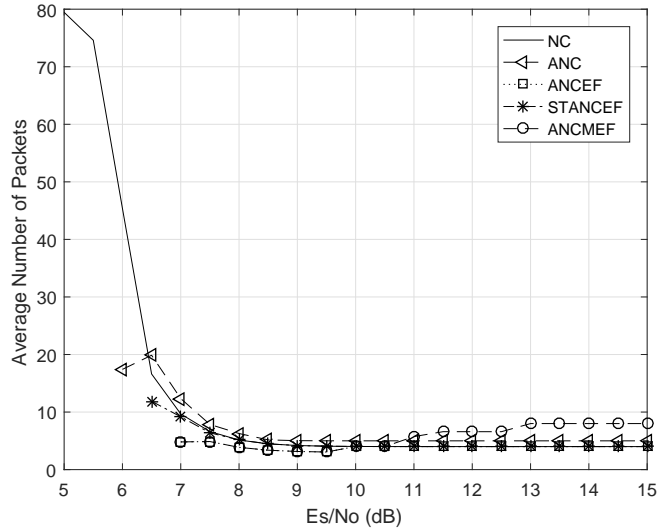

**Fig. 2.** Average number of sent packets for variable $E_s/N_0$ values and 10 $m/s$ mobility speed.

However, the maximum constrained batch size in ANCMEF is affected not only by $E_s/N_0$ but also by modulation method and energy consumption, so at low SNR it behaves similarly to the NC, then, going to higher SNR, it selects a higher modulation order allowing to increase the transmission reliability by the exploitation of higher diversity or modulation order. However, the increase of number of packets of ANCMEF is associated to shorter size packets that allows for equivalent energy per symbol for all modulation orders and across all schemes.

In Fig. 3, we can see the behavior in terms of transmission delay for the proposed schemes in a GEO satellite scenario. The proposed schemes have the higher delay for low SNR values; this is due to the fact that the energy efficient adaptive schemes adapt its transmission to small size batches at the low $E_s/N_0$ associated with high erasures. Therefore, the transmission suffers from extra waiting times for acknowledgment at the end of each short batch. Thus, as illustrated in Fig. 3, the time spent waiting for acknowledgment is very large compared to the time of delivering the coded packets. After a certain $E_s/N_0$ value, the schemes have similar performance due to the number of packets in each batch that has been increased and then converged to the exact dof value of the non-adaptive NC scheme. ANCMEF is not an exception since a normalization on the average number of packets with $\log m$ matches with the dof for such shorter length packets. It is indeed straightforward to understand the delay saturation of all schemes to its minimal value at the high SNR.

Fig. 4 presents the throughput for the measured schemes NC and ANC with the proposed ones, i.e., ANCEF, STANCEF and ANCMEF. For low and intermediate $E_s/N_0$ values, the schemes show remarkable differences but all of them give less throughput than ANC and NC schemes due to the reduced number of transmitted packets. However, we emphasize that the main aim of these schemes is to avoid any source of energy consumption that rise due to bad channel conditions, hence, our schemes favors to be silent from transmission, than consuming energy by limiting the transmitted packets, and its utilization to adaptive coding techniques allows for reliable transmission and less energy due to decreased re-transmissions encountered. Thus, achieving less throughput is understandable. Furthermore, its worth to observe that at a certain $E_s/N_0$, all the schemes converge to maximum saturation throughput independent of mobile speed or channel variation. While, ANCEF and ANCMEF almost coincide in the performance behavior but not the reliability, we can see that STANCEF tries to build a trade-off that resonates just in a limited throughput gain due to its dof one step loss tractability. In fact, such a small gain in throughput is shown to be associated with a small cost in the energy. For medium $E_s/N_0$ the extra complexity due to self tractability and excess transmissions has no gains, therefore, we see that STANCEF throughput performance deteriorate with respect to the ANCEF and ANCMEF. Finally, Fig. 5 presents the most important part of this study, the total energy consumption that has been calculated using both the spectral noise level as $N_0 = -107dBm$ and the expected time needed to send 4 coded packets/dofs for each $E_s/N_0$ of each scheme. It is possible to note that all the proposed schemes allow significant gains in terms of energy consumption with respect to the benchmarks. Its clear that at the low SNR we can see STANCEF gains roughly $15mWatt/Hz$ with respect to the NC scheme. Moreover, it is worth to see that ANCEF and ANCMEF reduce remarkably the energy consumption by gaining up to roughly $20mWatt/Hz$ with respect to the NC scheme; such an amount might seem to be small, however, this represents very high figure in a large scale system with multiple receivers. Furthermore, at the moderate $E_s/N_0$ we can see that due to the shorter sizes of the packets length used for higher modulation orders, the ANCMEF continues to have highest energy efficiency with further gains. Finally, at the high $E_s/N_0$ beyond $10dB$, is associated with an increase in the transmitted message which necessarily increases the overall system energy consumption.

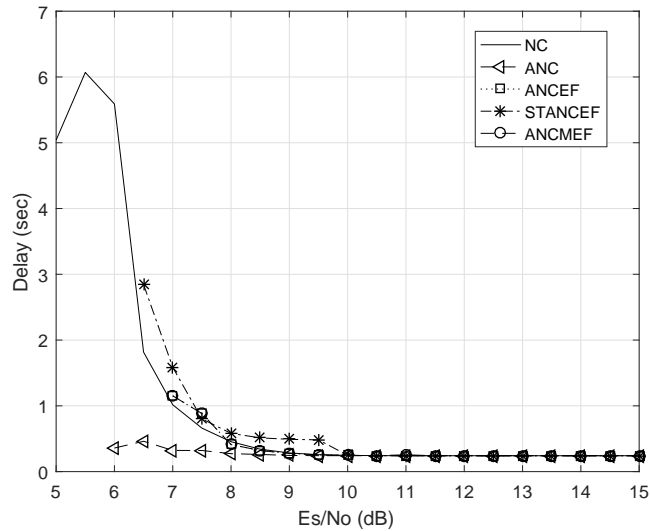

**Fig. 3.** Transmission Delay in a GEO Satellite scenario for variable $E_s/N_0$ values and $10\ m/s$ mobility speed.

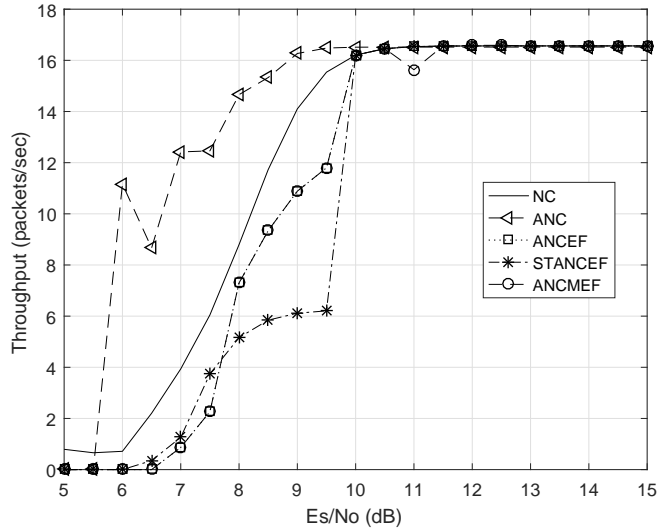

**Fig. 4.** Throughput in a GEO Satellite scenario for variable $E_s/N_0$ values and 10 $m/s$ mobility speed.

## 5 Conclusions

This paper addresses energy efficient adaptive network coding schemes for land satellite mobile with time varying channel. We proposed three novel adaptive physical layer aware schemes for coded packet transmission over LMS channel. Those schemes compensate for the lost degrees of freedom by tracking the packet erasures over time. The novelty of such schemes is expressed by their remarkable energy savings due to adaption to a set of various factors such as channel quality, that inherently adapts to the mobile speed, and thus allows due to smart silence/transmission periods to significant energy savings. Finally, we emphasize that, at SNR values high enough for reliable transmission, the schemes can be switched off to allow for a reduction in the processing power at the transmitter side.

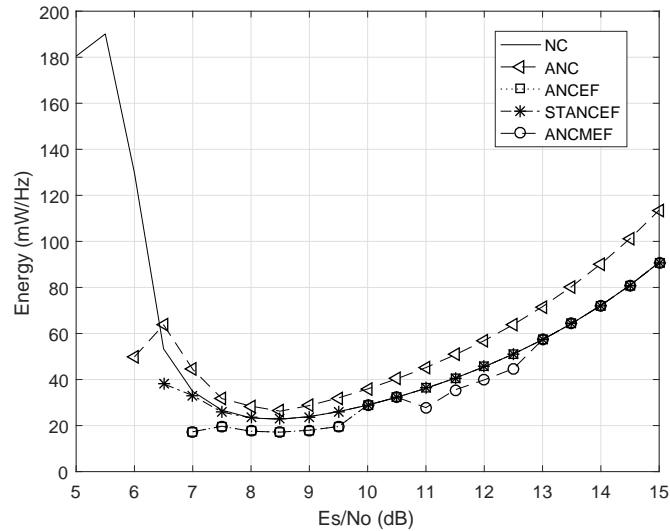

**Fig. 5.** Energy consumed in a GEO Satellite scenario for variable $E_s/N_0$ values and 10 $m/s$ mobility speed.